\documentclass[epj]{svjour}
\usepackage{graphics}
\newcommand{\bm}[1]{\mbox{\boldmath${#1}$}}

\def\inner#1#2{{\bm #1}\cdot {\bm #2}}

\def\Iprot#1#2{P^{#1#2}_{3/2}}
\def\Iproo#1#2{P^{#1#2}_{1/2}}
\def\Sprot#1#2{\Gamma^{#1#2}_{3/2}}
\def\Sproo#1#2{\Gamma^{#1#2}_{1/2}}
\def\half{\frac{1}{2}}

\begin{document}

\title{Baryon Tri-local Interpolating Fields}
\author{Hua-Xing Chen}
\institute{School of Physics and Nuclear Energy Engineering, Beihang University, Beijing 100191, China}
%
\date{Received: date / Revised version: date}
%
\abstract{
We systematically investigate tri-local (non-local) three-quark baryon fields with $U_L(2) \times U_R(2)$ chiral symmetry, according to their Lorentz and isospin (flavor) group representations. We note that they can also be called as ``nucleon wave functions'' due to this full non-locality. We study their chiral transformation properties and find all the possible chiral multiplets consisting of $J=\frac12$ and $J=\frac32$ baryon fields. We find that the axial coupling constant $|g_A| = \frac{5}{3}$ is only for nucleon fields belonging to the chiral representation $(\frac12,1) \oplus (1,\frac12)$, which contains both nucleon and $\Delta$ fields. Moreover, all the nucleon fields belonging to this representation have $|g_A| = \frac{5}{3}$.
\PACS{
      {11.30.Rd}{Chiral symmetries}   \and
      {12.38.-t}{Quantum chromodynamics}   \and
      {14.20.Gk}{Baryon resonances}
     } 
} 
\maketitle
%


\section{Introduction}
\label{intro}

The group theory has been applied to particle physics since its beginning. Particularly, the $SU(2)_L \otimes SU(2)_R$ and $SU(3)_L \otimes SU(3)_R$ chiral groups play important roles in low-energy hadron physics, and the relevant physics is called chiral symmetry~\cite{georgi,Ohnuki88,lee72,lee81,Slansky:1981yr,Weinberg:1969hw}. For five decades now, chiral symmetry and its spontaneous breaking have been studied using different methods because of their importance in elucidating the origin of hadron masses~\cite{chiral}. In Refs.~\cite{Weinberg:1969hw}, S.~Weinberg pointed out that the algebraic aspects of chiral symmetry are worth studying. Previous studies found some QCD identities for baryon currents (fields) using an algebraic method~\cite{identities}. Other scholars systematically investigated three-quark baryon fields with the $SU(2)_L \otimes SU(2)_R$ chiral symmetry, and we investigated those with the $SU(3)_L \otimes SU(3)_R$ chiral symmetry~\cite{Nagata:2007di,Chen:2008qv}.

Until now, most algebraic studies investigate local hadron fields. However, non-local ones should also be studied. They are important in the study of excited hadron states since (in the continuum limit) the local three-quark baryon fields can not have a spin larger than $3/2$~\cite{Ohnuki88}. To extend the algebraic studies on chiral symmetry, we systematically investigated bi-local three-quark baryon fields with the $SU(2)_L \otimes SU(2)_R$ chiral symmetry, where two quark fields are at the same place but the third one is at a different position~\cite{Dmitrasinovic:2011yf}. In this paper, we investigate tri-local (non-local) three-quark baryon fields, where three quarks are all at different positions. Due to this full non-locality they can also be called as ``nucleon wave functions'', which plays an important role in QCD process predictions. For example, the wave functions of $B$ and $D$ mesons are studied in order to investigate the mechanism of heavy meson decays~\cite{BDmesons}. The baryon (nucleon) wave functions are also studied in order to investigate the baryon structure as well as their decay properties~\cite{baryon}. 

In this paper, We systematically investigate these baryon fields according to their Lorentz and isospin (flavor) group representations. We study their chiral transformation properties and find all possible chiral multiplets consisting of $J=\frac12$ and $J=\frac32$ baryon fields. Our procedures mostly follow Ref.~\cite{Dmitrasinovic:2011yf}, but the calculations are a bit more complicated. The obtained baryon currents can be used in the lattice QCD and the QCD sum rule studies~\cite{sumrule}. We also study axial charges, which are important in understanding both electro-weak and strong interactions~\cite{axialcharge}.

This paper is organized as follows. In Sect.~\ref{sec:field} we systematically classify tri-local (non-local) three-quark baryon fields, according to their Lorentz and isospin (flavor) group representations. In Sect.~\ref{sec:transform} we study the behaviors of these fields under the Abelian $U(1)_A$ and non-Abelian $SU(2)_A$ chiral transformations. Finally, we summarize the obtained chiral multiplets in Sect.~\ref{sec:summary} with some discussions.

\section{Baryon Fields}
\label{sec:field}

A non-trivial interpolating field $B(x,y,z)$ coupling to baryons composed of three-quark fields can be generally written as
\begin{eqnarray}
B(x,y,z) &\sim& \epsilon_{abc} \left( q^{aT}_A (x) \Gamma_1 q^b_B(y)\right) \Gamma_2 q^c_C(z) \label{eq:baryonfield}
\end{eqnarray}
where $a,b,c$ denote the color and $A,B,C$ the flavor indices, $C = i\gamma_2 \gamma_0$ is the charge-conjugation operator. $q_A(x)=(u(x),\;d(x))^T$ is an iso-doublet quark field at location $x$, and the superscript $T$ represents the transpose of the Dirac indices only (the flavor and color indices are {\it not} transposed). The antisymmetric tensor in color space $\epsilon_{abc}$ ensures the baryons are in a color singlet state. Since it always exists, we shall omit it from now on. The matrices $\Gamma_{1,2}$ are Dirac matrices which describe the Lorentz structure. With a suitable choice of $\Gamma_{1,2}$ and taking a combination of indices of $A, B$ and $C$, the baryon fields are defined so that they form an irreducible representation of the Lorentz and flavor groups.

Previous studies investigated on local and bi-local baryon fields~\cite{Nagata:2007di,Dmitrasinovic:2011yf}. The present study investigates a more general case, the tri-local three-quark baryon fields, where the positions of quarks are all different ($x \neq y \neq z$). We note that due to this full non-locality we can also call them ``nucleon wave functions''. The procedures mostly follow our previous reference~\cite{Dmitrasinovic:2011yf}, but a bit more complicated. Basically, the Fierz transformation is used to change the positions of quark fields. Afterwhich, we can apply the Pauli principle to check the antisymmetric properties of every pair of quarks.

We classify tri-local three-quark baryon fields according to their spin and isospin, both of which can be either $1/2$ or $3/2$. To perform the classification and simplify the notations, a ``tilde-transposed'' quark field $\tilde{q}$ should be introduced:
\begin{eqnarray}
\tilde{q}=q^T C\gamma_5 (i\tau_2) \, ,
\end{eqnarray}
Moreover, we absorb the flavor indices into quark fields and write the baryon field as
\begin{eqnarray}
B(x,y,z) &\sim& \left( \tilde{q}^{T}(x) \tilde \Gamma_1 q(y)\right) \tilde \Gamma_2 q(z) \, . \label{eq:baryongeneral}
\end{eqnarray}
The matrices $\tilde \Gamma_{1,2}$ also contain some flavor structure.

Considering the presence of three different positions ($x \neq y \neq z$), the possible baryon fields are $B(x,y,z)$, $B(y,z,x)$, $B(z,x,y)$, $B(x,z,y)$, $B(z,y,x)$, and $B(y,x,z)$. However, we can simply change the positions of the first and second quark fields, for example,
\begin{eqnarray}\label{eq:relation}
q^{aT}_A(x) \gamma_5 q^b_B(y) = - q^{bT}_B(y) \gamma_5 q^a_A(x) \, .
\end{eqnarray}
Therefore, we only need to consider three of them. We find that $B(x,y,z)$, $B(y,z,x)$, and $B(z,x,y)$ can be related using only one transformation matrix. Thus, we use these three fields in the following subsections.

\subsection{$J = {1 \over 2}$ and $I = {1 \over 2}$}

In this subsection, we study the first case $D(\frac12,0)_{I=\frac12}$, where the representation of the Lorentz group is $D(\frac12,0)$ and the isospin is $I=\frac12$. We find 30 tri-local baryon fields of $J = {1 \over 2}$ and $I = {1 \over 2}$:
\begin{eqnarray}
&& \left. \begin{array}{l}
\left\{ \begin{array}{l} N_1 = ( \tilde{q}(x) q(y) ) q(z)\, , \\
N_2 = ( \tilde{q}(x) \gamma_5 q(y) ) \gamma_5 q(z)\, ,  \\
N_3 = ( \tilde{q}(x) \gamma_\mu q(y) ) \gamma^\mu q(z)\, ,  \\
N_4 = ( \tilde{q}(x) \gamma_\mu \gamma_5 \tau^i q(y) )
\gamma^\mu \gamma_5 \tau^i q(z)\, ,  \\
N_5 = ( \tilde{q}(x) \sigma_{\mu\nu} \tau^i q(y) ) \sigma^{\mu\nu}
\tau^i q(z)\, , \end{array} \right.
\\ \nonumber
\left\{ \begin{array}{l} N_{6} = ( \tilde{q}(x) \tau^i q(y) ) \tau^i q(z)\, , \\
N_{7} = ( \tilde{q}(x) \gamma_5 \tau^i q(y) ) \gamma_5 \tau^i q(z)\, , \\
N_{8} = ( \tilde{q}(x) \gamma_\mu \tau^i q(y) ) \gamma^\mu \tau^i q(z)\, , \\
N_{9} = ( \tilde{q}(x) \gamma_\mu \gamma_5 q(y) )
\gamma^\mu \gamma_5 q(z)\, , \\
N_{10} = ( \tilde{q}(x) \sigma_{\mu\nu} q(y) ) \sigma^{\mu\nu}
q(z)\, , \end{array} \right.
\end{array} \right. \\ \nonumber &&
\left. \begin{array}{l} \left\{ \begin{array}{l}
N_{11} = ( \tilde{q}(z) q(x) ) q(y)\, , \\
N_{12} = ( \tilde{q}(z) \gamma_5 q(x) ) \gamma_5 q(y)\, , \\
N_{13} = ( \tilde{q}(z) \gamma_\mu q(x) ) \gamma^\mu q(y)\, , \\
N_{14} = ( \tilde{q}(z) \gamma_\mu \gamma_5 \tau^i q(x) )
\gamma^\mu \gamma_5 \tau^i q(y)\, , \\
N_{15} = ( \tilde{q}(z) \sigma_{\mu\nu} \tau^i q(x) )
\sigma^{\mu\nu} \tau^i q(y)\, , \end{array} \right.
\\ \nonumber
\left\{ \begin{array}{l}
N_{16} = ( \tilde{q}(z) \tau^i q(x) ) \tau^i q(y)\, , \\
N_{17} = ( \tilde{q}(z) \gamma_5 \tau^i q(x) ) \tau^i \gamma_5 q(y)\, , \\
N_{18} = ( \tilde{q}(z) \gamma_\mu \tau^i q(x) ) \gamma^\mu \tau^i q(y)\, , \\
N_{19} = ( \tilde{q}(z) \gamma_\mu \gamma_5 q(x) )
\gamma^\mu \gamma_5 q(y)\, , \\
N_{20} = ( \tilde{q}(z) \sigma_{\mu\nu} q(x) ) \sigma^{\mu\nu}
q(y)\, , \end{array} \right.
\end{array} \right. \\ \nonumber &&
\left. \begin{array}{l} \left\{ \begin{array}{l}
N_{21} = ( \tilde{q}(y) q(z) ) q(x)\, , \\
N_{22} = ( \tilde{q}(y) \gamma_5 q(z) ) \gamma_5 q(x)\, , \\
N_{23} = ( \tilde{q}(y) \gamma_\mu q(z) ) \gamma^\mu q(x)\, , \\
N_{24} = ( \tilde{q}(y) \gamma_\mu \gamma_5 \tau^i q(z) )
\gamma^\mu \gamma_5 \tau^i q(x)\, , \\
N_{25} = ( \tilde{q}(y) \sigma_{\mu\nu} \tau^i q(z) )
\sigma^{\mu\nu} \tau^i q(x)\, , \end{array} \right.
\\ \nonumber
\left\{ \begin{array}{l}
N_{26} = ( \tilde{q}(y) \tau^i q(z) ) \tau^i q(x)\, , \\
N_{27} = ( \tilde{q}(y) \gamma_5 \tau^i q(z) ) \tau^i \gamma_5 q(x)\, , \\
N_{28} = ( \tilde{q}(y) \gamma_\mu \tau^i q(z) ) \gamma^\mu \tau^i q(x)\, , \\
N_{29} = ( \tilde{q}(y) \gamma_\mu \gamma_5 q(z) )
\gamma^\mu \gamma_5 q(x)\, , \\
N_{30} = ( \tilde{q}(y) \sigma_{\mu\nu} q(z) ) \sigma^{\mu\nu}
q(x)\, . \end{array} \right.
\end{array} \right.
\end{eqnarray}
The latter 20 fields $N_{11} \cdots N_{30}$ are the Fierz transformed fields of the former ten fields $N_{1} \cdots N_{10}$. Using the Fierz identities for the Dirac spin and isospin indices, we obtain the following identities:
\begin{eqnarray}
{\bf N}_A = \mathcal{A}^{{1}{1}} {\bf N}_B \, ,
{\bf N}_B = \mathcal{A}^{{1}{1}} {\bf N}_C \, ,
{\bf N}_C = \mathcal{A}^{{1}{1}} {\bf N}_A \, .
\end{eqnarray}
where ${\bf N_A}$, ${\bf N_B}$, and ${\bf N_C}$ denote the baryon vectors
\begin{eqnarray}
\nonumber {\bf N}_A &=& (N_1, N_2, \cdots , N_{10})^{\rm T} \, ,
\\ {\bf N}_B &=& (N_{11}, N_{12}, \cdots , N_{20})^{\rm T} \, ,
\\ \nonumber {\bf N}_C &=& (N_{21}, N_{22}, \cdots , N_{30})^{\rm T} \, ,
\end{eqnarray}
and $\mathcal{A}^{{1}{1}}$ is a $10\times10$ transformation matrix
\begin{eqnarray}
\mathcal{A}^{{1}{1}} = \frac18\left(\begin{array}{cccccccccc}
1 & 1 & 1 & -1 & \frac12 & -1 & -1 & -1 & 1 & -\frac12 \\
1 & 1 & -1 & 1 & \frac12 & -1 & -1 & 1 & -1 & -\frac12 \\
4 & -4 & -2 & -2 & 0 & -4 & 4 & 2 & 2 & 0 \\
-12 & 12 & -6 & 2 & 0 & -4 & 4 & -2 & 6 & 0 \\
36 & 36 & 0 & 0 & 2 & 12 & 12 & 0 & 0 & 6 \\
-3 & -3 & -3 & -1 & \frac12 & -1 & -1 & -1 & -3 & \frac32 \\
-3 & -3 & 3 & 1 & \frac12 & -1 & -1 & 1 & 3 & \frac32 \\
-12 & 12 & 6 & -2 & 0 & -4 & 4 & 2 & -6 & 0 \\
4 & -4 & 2 & 2 & 0 & -4 & 4 & -2 & -2 & 0 \\
-12 & -12 & 0 & 0 & 2 & 12 & 12 & 0 & 0 & -2
\end{array} \right ) \, .
\end{eqnarray}
This matrix is non-singular and it satisfies $(\mathcal{A}^{{1}{1}})^3 = 1$. The solution is
\begin{eqnarray}
{\bf N}_C = (\mathcal{A}^{{1}{1}})^{-1} {\bf N}_B = (\mathcal{A}^{{1}{1}})^{-2} {\bf N}_A \, .
\end{eqnarray}
Therefore, we obtain a ``trivial'' result that all the ten tri-local baryon fields $N_{1} \cdots N_{10}$ are non-zero as well as complete and independent. Other 20 fields $N_{11} \cdots N_{30}$ are related to these ten fields through the Fierz transformation.

\subsection{$J = {1 \over 2}$ and $I = {3 \over 2}$}

In this subsection, we study the case $D(\frac12,0)_{I=\frac32}$. We find $15$ bi-local baryon fields of $J = {1 \over 2}$ and $I = {3 \over 2}$:
\begin{eqnarray}\nonumber &&
\left. \begin{array}{l} \left\{\begin{array}{l} \Delta^i_4 = (
\tilde{q}(x) \gamma_\mu \gamma_5 \tau^j q(y) ) \gamma^\mu \gamma_5
P^{ij}_{3/2} q(z)\, , \\
\Delta^i_5 = ( \tilde{q}(x) \sigma_{\mu\nu} \tau^j q(y) )
\sigma^{\mu\nu} P^{ij}_{3/2} q(z)\, , \\
\end{array}\right.
\\ \nonumber
\left\{\begin{array}{l} \Delta^i_6 = ( \tilde{q}(x) \tau^j q(y)
) P^{ij}_{3/2} q(z)\, , \\
\Delta^i_7 = ( \tilde{q}(x) \gamma_5 \tau^j q(y) ) \gamma_5
P^{ij}_{3/2} q(z)\, , \\
\Delta^i_8 = ( \tilde{q}(x) \gamma_\mu \tau^j q(y) ) \gamma^\mu
P^{ij}_{3/2} q(z)\, ,
\end{array}\right.
\end{array}\right. \\ &&
\left. \begin{array}{l} \left\{\begin{array}{l} \Delta^i_{14} =
( \tilde{q}(z) \gamma_\mu \gamma_5 \tau^j q(x) ) \gamma^\mu \gamma_5 P^{ij}_{3/2} q(y)\, , \\
\Delta^i_{15} = ( \tilde{q}(z) \sigma_{\mu\nu} \tau^j q(x) ) \sigma^{\mu\nu} P^{ij}_{3/2} q(y)\, , \\
\end{array}\right.
\\ \nonumber
\left\{\begin{array}{l} \Delta^i_{16} = ( \tilde{q}(z)
\tau^j q(x) ) P^{ij}_{3/2} q(y)\, , \\
\Delta^i_{17} = ( \tilde{q}(z) \gamma_5 \tau^j q(x) ) \gamma_5 P^{ij}_{3/2} q(y)\, , \\
\Delta^i_{18} = ( \tilde{q}(z) \gamma_\mu \tau^j q(x) ) \gamma^\mu
P^{ij}_{3/2} q(y)\, ,
\end{array}\right.
\end{array}\right. \\ \nonumber &&
\left. \begin{array}{l} \left\{\begin{array}{l} \Delta^i_{24} =
( \tilde{q}(y) \gamma_\mu \gamma_5 \tau^j q(z) ) \gamma^\mu \gamma_5 P^{ij}_{3/2} q(x)\, , \\
\Delta^i_{25} = ( \tilde{q}(y) \sigma_{\mu\nu} \tau^j q(z) ) \sigma^{\mu\nu} P^{ij}_{3/2} q(x)\, , \\
\end{array}\right.
\\ \nonumber
\left\{\begin{array}{l} \Delta^i_{26} = ( \tilde{q}(y)
\tau^j q(z) ) P^{ij}_{3/2} q(x)\, , \\
\Delta^i_{27} = ( \tilde{q}(y) \gamma_5 \tau^j q(z) ) \gamma_5 P^{ij}_{3/2} q(x)\, , \\
\Delta^i_{28} = ( \tilde{q}(y) \gamma_\mu \tau^j q(z) ) \gamma^\mu
P^{ij}_{3/2} q(x)\, .
\end{array}\right.
\end{array}\right.
\end{eqnarray}
The latter ten fields $\Delta_{14} \cdots \Delta_{18}$ and $\Delta_{24} \cdots \Delta_{28}$ are the Fierz transformed fields of the former five fields $\Delta_{4} \cdots \Delta_{8}$. Here, $\Iprot{i}{j}$ is the isospin-projection operator for $I=\frac32$. We define it together with the isospin-projection operator $\Iproo{i}{j}$ for $I=\frac12$:
\begin{eqnarray}
\Iprot{i}{j}=\delta^{ij}-\frac13 \tau^i\tau^j \, ,
\Iproo{i}{j}=\frac13 \tau^i\tau^j \, ,
\end{eqnarray}
which satisfy
\begin{eqnarray}
\tau^i P^{ij}_{\frac32}=0 \, .
\end{eqnarray}
Using the Fierz identities for the Dirac spin and isospin indices, we obtain the following identities:
\begin{eqnarray}
{\bf \Delta}^i_A = \mathcal{A}^{{1}{3}} {\bf \Delta}^i_B \, ,
{\bf \Delta}^i_B = \mathcal{A}^{{1}{3}} {\bf \Delta}^i_C \, ,
{\bf \Delta}^i_C = \mathcal{A}^{{1}{3}} {\bf \Delta}^i_A \, .
\end{eqnarray}
where
\begin{eqnarray}\nonumber
{\bf \Delta}^i_A &=& ( \Delta^i_6 \, , \Delta^i_7 \, , \Delta^i_8 \,
, \Delta^i_4 \, , \Delta^i_5 )^{\rm T} \, ,
\\
{\bf \Delta}^i_B &=& ( \Delta^i_{16} \, , \Delta^i_{17} \, ,
\Delta^i_{18} \, , \Delta^i_{14} \, , \Delta^i_{15} )^{\rm T} \, ,
\\ \nonumber
{\bf \Delta}_C^i &=& ( \Delta^i_{26} \, , \Delta^i_{27} \, ,
\Delta^i_{28} \, , \Delta^i_{24} \, , \Delta^i_{25} )^{\rm T} \, ,
\end{eqnarray}
and $\mathcal{A}^{{1}{3}}$ is a $5\times5$ transformation matrix
\begin{eqnarray}
\mathcal{A}^{{1}{3}} = \frac14\left(\begin{array}{ccccc}
1 & 1 & 1 & 1 & -\frac12 \\
1 & 1 & -1 & -1 & -\frac12 \\
4 & -4 & -2 & 2 & 0 \\
4 & -4 & 2 & -2 & 0 \\
-12 & -12 & 0 & 0 & -2
\end{array} \right) \, .
\end{eqnarray}
Again, this matrix is non-singular and it satisfies $(\mathcal{A}^{{1}{3}})^3 = 1$. The solution is
\begin{eqnarray}
{\bf \Delta}_C^i = (\mathcal{A}^{{1}{3}})^{-1} {\bf \Delta}_B^i = (\mathcal{A}^{{1}{3}})^{-2} {\bf \Delta}_A^i \, .
\end{eqnarray}
Therefore, five fields are complete and independent.

\subsection{$J = {3 \over 2}$ and $I = {1 \over 2}$}

The three-quark baryon field of $J = {3 \over 2}$ contains either one free Lorentz index ($N_\mu$) or two antisymmetric Lorentz indices ($N_{\mu \nu} = - N_{\nu \mu}$). The former case has a Lorentz representation $D(1,\frac12)_{I=\frac12}$, and we find 18 tri-local baryon fields
\begin{eqnarray}\nonumber &&
\left. \begin{array}{l} \left\{ \begin{array}{l} N_{3\mu} = (
\tilde{q}(x) \gamma_\nu q(y) ) \Gamma^{\mu\nu}_{3/2} \gamma_5 q(z)\, , \\
N_{4\mu} = ( \tilde{q}(x) \gamma_\nu \gamma_5 \tau^i q(y) )
\Gamma^{\mu\nu}_{3/2} \tau^i q(z)\, , \\
N_{5\mu} = ( \tilde{q}(x) \sigma_{\alpha\beta} \tau^i q(y) )
\Gamma^{\mu\alpha}_{3/2} \gamma^\beta \gamma_5 \tau^i q(z)\, ,
\end{array}\right.
\\ \nonumber
\left\{ \begin{array}{l} N_{8\mu} = (
\tilde{q}(x) \gamma_\nu
\tau^i q(y) ) \Gamma^{\mu\nu}_{3/2} \gamma_5 \tau^i q(z)\, , \\
N_{9\mu} = ( \tilde{q}(x) \gamma_\nu \gamma_5 q(y) )
\Gamma^{\mu\nu}_{3/2} q(z)\, , \\
N_{10\mu} = ( \tilde{q}(x) \sigma_{\alpha\beta} q(y) )
\Gamma^{\mu\alpha}_{3/2} \gamma^\beta \gamma_5 q(z)\, ,
\end{array}\right.
\end{array}\right. \\ &&
\left. \begin{array}{l} \left\{ \begin{array}{l} N_{13\mu} = (
\tilde{q}(z) \gamma_\nu q(x) ) \Gamma^{\mu\nu}_{3/2}
\gamma_5 q(y)\, , \\
N_{14\mu} =
( \tilde{q}(z) \gamma_\nu \gamma_5 \tau^i q(x) ) \Gamma^{\mu\nu}_{3/2} \tau^i q(y)\, , \\
N_{15\mu} = ( \tilde{q}(z) \sigma_{\alpha\beta} \tau^i q(x)
)\Gamma^{\mu\alpha}_{3/2} \gamma^\beta \gamma_5 \tau^i q(y)\, ,
\end{array}\right.
\\ \nonumber
\left\{ \begin{array}{l} N_{18\mu} = (
\tilde{q}(z) \gamma_\nu \tau^i q(x) ) \Gamma^{\mu\nu}_{3/2}
\gamma_5 \tau^i q(y)\, , \\
N_{19\mu} =
( \tilde{q}(z) \gamma_\nu \gamma_5 q(x) ) \Gamma^{\mu\nu}_{3/2} q(y)\, , \\
N_{20\mu} = ( \tilde{q}(z) \sigma_{\alpha\beta} q(x)
)\Gamma^{\mu\alpha}_{3/2} \gamma^\beta \gamma_5 q(y)\, ,
\end{array}\right.
\end{array}\right. \\ \nonumber &&
\left. \begin{array}{l} \left\{ \begin{array}{l} N_{23\mu} = (
\tilde{q}(y) \gamma_\nu q(z) ) \Gamma^{\mu\nu}_{3/2}
\gamma_5 q(x)\, , \\
N_{24\mu} =
( \tilde{q}(y) \gamma_\nu \gamma_5 \tau^i q(z) ) \Gamma^{\mu\nu}_{3/2} \tau^i q(x)\, , \\
N_{25\mu} = ( \tilde{q}(y) \sigma_{\alpha\beta} \tau^i q(z)
)\Gamma^{\mu\alpha}_{3/2} \gamma^\beta \gamma_5 \tau^i q(x)\, ,
\end{array}\right.
\\ \nonumber
\left\{ \begin{array}{l} N_{28\mu} = (
\tilde{q}(y) \gamma_\nu \tau^i q(z) ) \Gamma^{\mu\nu}_{3/2}
\gamma_5 \tau^i q(x)\, , \\
N_{29\mu} =
( \tilde{q}(y) \gamma_\nu \gamma_5 q(z) ) \Gamma^{\mu\nu}_{3/2} q(x)\, , \\
N_{30\mu} = ( \tilde{q}(y) \sigma_{\alpha\beta} q(z)
)\Gamma^{\mu\alpha}_{3/2} \gamma^\beta \gamma_5 q(x)\, .
\end{array}\right.
\end{array}\right.
\end{eqnarray}
The latter 12 fields are the Fierz transformed fields of the former six fields $N_{3 \mu} \cdots N_{5 \mu}$ and $N_{8 \mu} \cdots N_{10 \mu}$. Similar to the isospin projection operators, we define $\Sprot{\mu}{\nu}$ and $\Sproo{\mu}{\nu}$, which are the spin-projection operators for the $J=\frac32$ and $J=\frac12$ states:
\begin{eqnarray}
\Sprot{\mu}{\nu}=g^{\mu\nu}-\frac14 \gamma^\mu\gamma^\nu \, ,
\Sproo{\mu}{\nu}=\frac14 \gamma^\mu\gamma^\nu \, . \label{10jan08eq1}
\end{eqnarray}
They satisfy
\begin{eqnarray}
\gamma_\mu \Sprot{\mu}{\nu} = 0 \, .
\end{eqnarray}
Using the Fierz identities for the Dirac spin and isospin indices, we obtain the following identities:
\begin{eqnarray}
{\bf N}_{A\mu} = \mathcal{A}^{{3}{1}} {\bf N}_{B\mu} \, ,
{\bf N}_{B\mu} = \mathcal{A}^{{3}{1}} {\bf N}_{C\mu} \, ,
{\bf N}_{C\mu} = \mathcal{A}^{{3}{1}} {\bf N}_{A\mu} \, .
\end{eqnarray}
where
\begin{eqnarray}\nonumber
{\bf N}_{A\mu} &=& ( N_{3\mu} \, , N_{4\mu} \, , N_{5\mu} \, , N_{8\mu} \, , N_{9\mu} \, , N_{10\mu} )^{\rm T} \, ,
\\
{\bf N}_{B\mu} &=& ( N_{13\mu} \, , N_{14\mu} \, , N_{15\mu} \, , N_{18\mu} \, , N_{19\mu} \, , N_{20\mu} )^{\rm T} \, ,
\\ \nonumber
{\bf N}_{C\mu} &=& ( N_{23\mu} \, , N_{24\mu} \, , N_{25\mu} \, , N_{28\mu} \, , N_{29\mu} \, , N_{30\mu} )^{\rm T} \, ,
\end{eqnarray}
and $\mathcal{A}^{{3}{1}}$ is a $6\times6$ transformation matrix
\begin{eqnarray}
\mathcal{A}^{{3}{1}} = \frac14\left(\begin{array}{cccccc}
1 & 1 & 1 & -1 & -1 & -1 \\
3 & -1 & 1 & 1 & -3 & 3 \\
6 & 2 & 0 & 2 & 6 & 0 \\
-3 & 1 & 1 & -1 & 3 & 3 \\
-1 & -1 & 1 & 1 & 1 & -1 \\
-2 & 2 & 0 & 2 & -2 & 0
\end{array}
\right) \, .
\end{eqnarray}
This matrix is non-singular and it satisfies $(\mathcal{A}^{{3}{1}})^3 = 1$. The solution is
\begin{eqnarray}
{\bf N}_{C\mu} = (\mathcal{A}^{{3}{1}})^{-1} {\bf N}_{B\mu} = (\mathcal{A}^{{3}{1}})^{-2} {\bf N}_{A\mu} \, .
\end{eqnarray}
Therefore, six fields are complete and independent.

For two antisymmetric Lorentz indices with the Lorentz representation $D(\frac32,0)_{I=\frac12}$, we find six fields:
\begin{eqnarray}
\nonumber && \left \{ \begin{array}{l}
N_{5\mu\nu} = ( \tilde{q}(x) \sigma_{\alpha\beta} \tau^i q(y) ) \Gamma^{\mu\nu\alpha\beta}_{3/2} \tau^i q(z)\, ,
\\ N_{10\mu\nu} = ( \tilde{q}(x) \sigma_{\alpha\beta} q(y) ) \Gamma^{\mu\nu\alpha\beta}_{3/2} q(z)\, ,
\end{array} \right .
\\ &&
\left \{ \begin{array}{l}
N_{15\mu\nu} = ( \tilde{q}(z) \sigma_{\alpha\beta} \tau^i q(x) ) \Gamma^{\mu\nu\alpha\beta}_{3/2} \tau^i q(y)\, ,
\\ N_{20\mu\nu} = ( \tilde{q}(z) \sigma_{\alpha\beta} q(x) ) \Gamma^{\mu\nu\alpha\beta}_{3/2} q(y)\, ,
\end{array} \right .
\\ \nonumber &&
\left \{ \begin{array}{l}
N_{25\mu\nu} = ( \tilde{q}(y) \sigma_{\alpha\beta} \tau^i q(z) ) \Gamma^{\mu\nu\alpha\beta}_{3/2} \tau^i q(x)\, ,
\\ N_{30\mu\nu} = ( \tilde{q}(y) \sigma_{\alpha\beta} q(z) ) \Gamma^{\mu\nu\alpha\beta}_{3/2} q(x)\, .
\end{array} \right .
\end{eqnarray}
The latter four fields are the Fierz transformed fields of the former two fields $N_{5 \mu \nu}$ and $N_{10 \mu \nu}$. Here, $\Gamma^{\mu\nu\alpha\beta}$ is another $J=\frac32$ projection operator defined as
\begin{eqnarray}
\Gamma^{\mu\nu\alpha\beta}=\left(g^{\mu\alpha}g^{\nu\beta} -\half g^{\nu\beta}\gamma^\mu\gamma^\alpha +\half g^{\mu\beta}\gamma^\nu\gamma^\alpha +\frac16 \sigma^{\mu\nu}\sigma^{\alpha\beta}\right)\, . \nonumber \\
\end{eqnarray}
Using the Fierz identities for the Dirac spin and isospin indices, we obtain the following identities:
\begin{eqnarray}
{\bf N}_{A\mu\nu} = \mathcal{B}^{{3}{1}} {\bf N}_{B\mu\nu} \, ,
{\bf N}_{B\mu\nu} = \mathcal{B}^{{3}{1}} {\bf N}_{C\mu\nu} \, ,
{\bf N}_{C\mu\nu} = \mathcal{B}^{{3}{1}} {\bf N}_{A\mu\nu} \, .
\nonumber \\
\end{eqnarray}
where
\begin{eqnarray}\nonumber
{\bf N}_{A\mu\nu} &=& ( N_{5\mu\nu} \, , N_{10\mu\nu} )^{\rm T} \, ,
\\
{\bf N}_{B\mu\nu} &=& ( N_{15\mu\nu} \, , N_{20\mu\nu} )^{\rm T} \, ,
\\ \nonumber
{\bf N}_{C\mu\nu} &=& ( N_{25\mu\nu} \, , N_{30\mu\nu} )^{\rm T} \, ,
\end{eqnarray}
and $\mathcal{B}^{{3}{1}}$ is a $2\times2$ transformation matrix
\begin{eqnarray}
\mathcal{B}^{{3}{1}} = \frac12 \left(\begin{array}{cc}
1 & -1 \\
-3 & -1
\end{array}
\right)
\end{eqnarray}
This matrix is non-singular and it satisfies $(\mathcal{B}^{{3}{1}})^3 = 1$. The solution is
\begin{eqnarray}
{\bf N}_{C\mu\nu} = (\mathcal{B}^{{3}{1}})^{-1} {\bf N}_{B\mu\nu} = (\mathcal{B}^{{3}{1}})^{-2} {\bf N}_{A\mu\nu} \, .
\end{eqnarray}
Therefore, two fields are complete and independent.

\subsection{$J = {3 \over 2}$ and $I = {3 \over 2}$}

In this subsection, we also need to consider the case of one Lorentz index ($\Delta^i_\mu$) and two antisymmetric Lorentz indices ($\Delta^i_{\mu\nu} = - \Delta^i_{\nu\mu}$). For the former one with the Lorentz representation $D(1,\frac12)_{I=\frac32}$, we find nine baryon fields
\begin{eqnarray}
\nonumber && \left\{\begin{array}{l}
\Delta^i_{4\mu} = ( \tilde{q}(x) \gamma_\nu \gamma_5 \tau^j q(y) ) \Gamma^{\mu\nu}_{3/2} P^{ij}_{3/2} q(z)\, ,
\\
\Delta^i_{5\mu} = ( \tilde{q}(x) \sigma_{\alpha\beta} \tau^j q(y) ) \Gamma^{\mu\alpha}_{3/2} \gamma^\beta \gamma_5 P^{ij}_{3/2} q(z)\, ,
\\
\Delta^i_{8\mu} = ( \tilde{q}(x) \gamma_\nu \tau^j q(y) ) \Gamma^{\mu\nu}_{3/2} \gamma_5 P^{ij}_{3/2} q(z)\, ,
\end{array}\right.
\\ && \left\{\begin{array}{l}
\Delta^i_{14\mu} = ( \tilde{q}(z) \gamma_\nu \gamma_5 \tau^j q(x) ) \Gamma^{\mu\nu}_{3/2} P^{ij}_{3/2} q(y)\, ,
\\
\Delta^i_{15\mu} = ( \tilde{q}(z) \sigma_{\alpha\beta} \tau^j q(x) )\Gamma^{\mu\alpha}_{3/2} \gamma^\beta \gamma_5 P^{ij}_{3/2} q(y)\, ,
\\
\Delta^i_{18\mu} = ( \tilde{q}(z) \gamma_\nu \tau^i q(x) ) \Gamma^{\mu\nu}_{3/2} \gamma_5 P^{ij}_{3/2} q(y)\, ,
\end{array}\right.
\\ \nonumber && \left\{\begin{array}{l}
\Delta^i_{24\mu} = ( \tilde{q}(y) \gamma_\nu \gamma_5 \tau^j q(z) ) \Gamma^{\mu\nu}_{3/2} P^{ij}_{3/2} q(x)\, ,
\\
\Delta^i_{25\mu} = ( \tilde{q}(y) \sigma_{\alpha\beta} \tau^j q(z) )\Gamma^{\mu\alpha}_{3/2} \gamma^\beta \gamma_5 P^{ij}_{3/2} q(x)\, ,
\\
\Delta^i_{28\mu} = ( \tilde{q}(y) \gamma_\nu \tau^i q(z) ) \Gamma^{\mu\nu}_{3/2} \gamma_5 P^{ij}_{3/2} q(x)\, .
\end{array}\right.
\end{eqnarray}
The latter six fields are the Fierz transformed fields of the former three fields $\Delta^i_{4 \mu}$, $\Delta^i_{5 \mu}$, and $\Delta^i_{8 \mu}$. Using the Fierz identities for the Dirac spin and isospin indices, we obtain the following identities:
\begin{eqnarray}
{\bf \Delta}^i_{A\mu} = \mathcal{A}^{{3}{3}} {\bf \Delta}^i_{B\mu} \, ,
{\bf \Delta}^i_{B\mu} = \mathcal{A}^{{3}{3}} {\bf \Delta}^i_{C\mu} \, ,
{\bf \Delta}^i_{C\mu} = \mathcal{A}^{{3}{3}} {\bf \Delta}^i_{A\mu} \, .
\nonumber \\
\end{eqnarray}
where
\begin{eqnarray}\nonumber
{\bf \Delta}^i_{A\mu} &=& ( \Delta^i_{8\mu} \, , \Delta^i_{4\mu} \, , \Delta^i_{5\mu} )^{\rm T} \, ,
\\
{\bf \Delta}^i_{B\mu} &=& ( \Delta^i_{18\mu} \, , \Delta^i_{14\mu} \, , \Delta^i_{15\mu} )^{\rm T} \, ,
\\ \nonumber
{\bf \Delta}^i_{C\mu} &=& ( \Delta^i_{28\mu} \, , \Delta^i_{24\mu} \, , \Delta^i_{25\mu} )^{\rm T} \, ,
\end{eqnarray}
and $\mathcal{A}^{{3}{3}}$ is a $3\times3$ transformation matrix
\begin{eqnarray}
\mathcal{A}^{{3}{3}} = - \frac12\left(\begin{array}{cccccc}
1 & -1 & -1 \\
-1 & 1 & -1 \\
-2 & -2 & 0
\end{array}
\right) \, .
\end{eqnarray}
This matrix is non-singular and it satisfies $(\mathcal{A}^{{3}{3}})^3 = 1$. The solution is
\begin{eqnarray}
{\bf \Delta}^i_{C\mu} = (\mathcal{A}^{{3}{3}})^{-1} {\bf \Delta}^i_{B\mu} = (\mathcal{A}^{{3}{3}})^{-2} {\bf \Delta}^i_{A\mu} \, .
\end{eqnarray}
Therefore, three fields are complete and independent.

Finally, for two antisymmetric Lorentz indices with the Lorentz representation $D(\frac32,0)_{I=\frac32}$, we find three $\Delta$ fields
\begin{eqnarray} \nonumber
\Delta^i_{5\mu\nu} &=& ( \tilde{q}(x) \sigma_{\alpha\beta} \tau^j
q(y) ) \Gamma^{\mu\nu\alpha\beta}_{3/2} P^{ij}_{3/2} q(z) \, ,
\\ \Delta^i_{15\mu\nu} &=& (
\tilde{q}(z) \sigma_{\alpha\beta} \tau^j q(x) )
\Gamma^{\mu\nu\alpha\beta}_{3/2} P^{ij}_{3/2} q(y) \, ,
\\ \nonumber \Delta^i_{25\mu\nu} &=& (
\tilde{q}(y) \sigma_{\alpha\beta} \tau^j q(z) )
\Gamma^{\mu\nu\alpha\beta}_{3/2} P^{ij}_{3/2} q(x) \, .
\end{eqnarray}
The latter two fields are the Fierz transformed fields of the first one $\Delta^i_{5 \mu \nu}$. Using the Fierz identities for the Dirac spin and isospin indices, we obtain the following identities:
\begin{eqnarray}\nonumber
\Delta^i_{5\mu\nu} = \Delta^i_{15\mu\nu} = \Delta^i_{25\mu\nu} \, .
\end{eqnarray}
Therefore, there is only one field, and itself is a complete basis.

\section{Chiral Transformations}
\label{sec:transform}

In this section, we study the chiral transformations of tri-local baryon fields. The analysis and notations in this section are similar to our previous paper~\cite{Dmitrasinovic:2011yf}, and the results are listed here. We do this procedure according to their spin in the following subsections.

\subsection{$J = {1 \over 2}$}

Under the Abelian $U(1)_A$ chiral transformation, the $D(\frac12,0)_{I=\frac12}$ three-quark baryon fields transform as
\begin{eqnarray}
&& \left \{
\begin{array}{l}
\delta_5 N_{1} = i a \gamma_5 (N_{1} + 2 N_{2}) \, ,
\\ \delta_5 N_{2} = i a \gamma_5 (2 N_{1} + N_{2}) \, ,
\\ \delta_5 N_{3} = - i a \gamma_5 N_{3} \, ,
\\ \delta_5 N_{4} = - i a \gamma_5 N_{4} \, ,
\\ \delta_5 N_{5} = 3 a \gamma_5 N_{5} \, ,
\end{array} \right .
\\ \nonumber &&
\left \{
\begin{array}{l}
\delta_5 N_{6} = i a \gamma_5 (N_{6} + 2 N_{7}) \, ,
\\ \delta_5 N_{7} = i a \gamma_5 (2 N_{6} + N_{7}) \, ,
\\ \delta_5 N_{8} = - i a \gamma_5 N_{8} \, ,
\\ \delta_5 N_{9} = - i a \gamma_5 N_{9} \, ,
\\ \delta_5 N_{10} = 3 a \gamma_5 N_{10} \, ,
\end{array} \right .
\end{eqnarray}
and the $D(\frac12,0)_{I=\frac32}$ baryon fields transform as
\begin{eqnarray}
&& \left \{ \begin{array}{l}
\delta_5 \Delta^i_{4} = - i a \gamma_5 \Delta^i_{4} \, ,
\\ \delta_5 \Delta^i_{5} = 3 i a \gamma_5 \Delta^i_{5} \, ,
\end{array} \right.
\\ \nonumber &&
\left \{ \begin{array}{l}
\delta_5 \Delta^i_{6} = i a \gamma_5 (\Delta^i_{6} + 2 \Delta^i_{7}) \, ,
\\ \delta_5 \Delta^i_{7} = i a \gamma_5 (2 \Delta^i_{6} + \Delta^i_{7}) \, ,
\\ \delta_5 \Delta^i_{8} = - i a \gamma_5 \Delta^i_{8} \, .
\end{array} \right.
\end{eqnarray}
We find that $N_1$ and $N_2$ can be reduced to irreducible components by taking the antisymmetric linear combination of the two nucleon fields:
\begin{eqnarray}
\delta_5 (N_{1} + N_{2}) &=& 3 i a \gamma_5 (N_{1} + N_{2}) \, ,
\\ \nonumber
\delta_5 (N_{1} - N_{2}) &=& - i a \gamma_5 (N_{1} - N_{2}) \, .
\end{eqnarray}
The same can be done for $N_{6}$, $N_{7}$, $\Delta^i_{6}$, and $\Delta^i_{7}$:
\begin{eqnarray}
\nonumber \delta_5 (N_{6} + N_{7}) &=& 3 i a \gamma_5 (N_{6} + N_{7}) \, ,
\\ \delta_5 (N_{6} - N_{7}) &=& - i a \gamma_5 (N_{6} - N_{7}) \, ,
\\ \nonumber \delta_5 (\Delta^i_{6} + \Delta^i_{7}) &=& 3 i a \gamma_5 (\Delta^i_{6} + \Delta^i_{7}) \, ,
\\ \nonumber \delta_5 (\Delta^i_{6} - \Delta^i_{7}) &=& - i a \gamma_5 (\Delta^i_{6} - \Delta^i_{7}) \, .
\end{eqnarray}

Under the non-Abelian $SU(2)_A$ chiral transformation, the $D(\frac12,0)_{I=\frac12}$ baryon fields transform as
\begin{eqnarray}
&& \nonumber \left \{ \begin{array}{l}
\delta_5^{\vec{a}} N_{1} = i \inner{a}{\tau} \gamma_5 N_{1} \, ,
\\ \delta_5 N_{2} = i \inner{a}{\tau} \gamma_5 N_{2} \, ,
\\ \delta_5^{\vec{a}} {N_3} = - i \inner{a}{\tau}\gamma_5 {N_3} - \frac23 i \inner{a}{\tau} \gamma_5 N_4 - 2i \gamma_5 \inner{a}{\Delta_4} \, ,
\\ \delta_5^{\vec{a}} N_4 = - 2 i \inner{a}{\tau}\gamma_5 {N_3} + \frac13 i \inner{a}{\tau}\gamma_5 N_4 - 2 i \gamma_5 \inner{a}{\Delta_4} \, ,
\\ \delta_{5}^{\vec{a}} N_5 = i \inner{a}{\tau} \gamma_5 N_5 \, ,
\end{array} \right .
\\ && \left \{ \begin{array}{l}
\delta_{5}^{\vec{a}} N_6 = - \frac13 i \inner{a}{\tau} \gamma_5 N_6 + \frac43 i \inner{a}{\tau} \gamma_5 N_7
\\ ~~~~~~~~~~ + 2 i \gamma_5 \inner{a}{\Delta_6} - 2 i \gamma_5 \inner{a}{\Delta_7} \, ,
\\ \delta_{5}^{\vec{a}} N_7 = - \frac13 i \inner{a}{\tau} \gamma_5 N_7 + \frac43 i \inner{a}{\tau} \gamma_5 N_6
\\ ~~~~~~~~~~ + 2 i \gamma_5 \inner{a}{\Delta_7} - 2 i \gamma_5 \inner{a}{\Delta_6} \, ,
\\ \delta_{5}^{\vec{a}} N_8 = \frac13 i \inner{a}{\tau} \gamma_5 N_8 - 2 i \inner{a}{\tau} \gamma_5 N_9 - 2 i \gamma_5 \inner{a}{\Delta_8} \, ,
\\ \delta_{5}^{\vec{a}} N_9 = - \frac23 i \inner{a}{\tau} \gamma_5 N_8 - i \inner{a}{\tau} \gamma_5 N_9 - 2 i \gamma_5 \inner{a}{\Delta_8} \, ,
\\ \delta_{5}^{\vec{a}} N_{10} = i \inner{a}{\tau} \gamma_5 N_{10} \, ,
\end{array} \right .
\end{eqnarray}
and the $D(\frac12,0)_{I=\frac32}$ baryon fields transform as
\begin{eqnarray}
&& \left \{ \begin{array}{l}
\delta_5^{\vec{a}} \Delta_4^i = - 2 i \gamma_5 a^j \Iprot{i}{j} {N_3} -\frac23 i\gamma_5 a^j P^{ij}_{3/2} N_4
\\ ~~~~~~~~~~ + \frac23 i \tau^i \gamma_5 \inner{a}{\Delta_4} - i\inner{a}{\tau}\gamma_5 \Delta_4^i \, ,
\\ \delta_5^{\vec{a}} \Delta_5^i = - 2 i \tau^i \gamma_5 \inner{a}{\Delta_5} + 3 i \inner{a}{\tau} \gamma_5 \Delta_5^i \, ,
\end{array} \right .
\\ \nonumber && \left \{ \begin{array}{l}
\delta_5^{\vec{a}} \Delta_6^i = - \frac43 i \tau^i \gamma_5 \inner{a}{\Delta_7} + 2 i \inner{a}{\tau} \gamma_5 \Delta_7^i -\frac23 i\gamma_5 a^j P^{ij}_{3/2} N_7
\\ ~~~~~~~~~~ - \frac23 i \tau^i \gamma_5 \inner{a}{\Delta_6} + i \inner{a}{\tau} \gamma_5 \Delta_6^i + \frac23 i\gamma_5 a^j P^{ij}_{3/2} N_6 \, ,
\\ \delta_5^{\vec{a}} \Delta_7^i = - \frac43 i \tau^i \gamma_5 \inner{a}{\Delta_6} + 2 i \inner{a}{\tau} \gamma_5 \Delta_6^i -\frac23 i\gamma_5 a^j P^{ij}_{3/2} N_6
\\ ~~~~~~~~~~ - \frac23 i \tau^i \gamma_5 \inner{a}{\Delta_7} + i \inner{a}{\tau} \gamma_5 \Delta_7^i + \frac23 i\gamma_5 a^j P^{ij}_{3/2} N_7 \, ,
\\ \delta_5^{\vec{a}} \Delta_8^i = - 2 i \gamma_5 a^j \Iprot{i}{j} {N_9} -\frac23 i\gamma_5 a^j P^{ij}_{3/2} N_8
\\ ~~~~~~~~~~ + \frac23 i \tau^i \gamma_5 \inner{a}{\Delta_8} - i\inner{a}{\tau}\gamma_5 \Delta_8^i \, .
\end{array} \right .
\end{eqnarray}
We find that $N_3$, $N_4$, and $\Delta_4^i$ can be reduced to irreducible components by taking the following linear combinations:
\begin{eqnarray}
&& \delta_5^{\vec{a}} ({N_3} - N_4) = i \inner{a}{\tau} \gamma_5 ({N_3} - N_4) \, ,
\\ \nonumber && \left \{ \begin{array}{l}
\delta_5^{\vec{a}}(3 {N_3} + N_4) = - i \gamma_5 \left[\frac53 \inner{a}{\tau} (3 {N_3} + N_4) + 8 \inner{a}{\Delta_4} \right] \, ,
\\ \delta_5^{\vec{a}} \Delta_4^i = - i\gamma_5 \left[\frac23 a^j P^{ij}_{3/2} (3 {N_3} + N_4) - \frac23 \tau^i \inner{a}{\Delta_4} + \inner{a}{\tau} \Delta_4^i \right] \, .
\end{array} \right .
\end{eqnarray}
The same can be done for $N_6$, $N_7$, $\Delta_6^i$, and $\Delta_7^i$:
\begin{eqnarray}
&& \delta_{5}^{\vec{a}} (N_6 + N_7) = i \inner{a}{\tau} \gamma_5 (N_6 + N_7) \, ,
\\ \nonumber && \left \{ \begin{array}{l}
\delta_{5}^{\vec{a}} (N_6 - N_7) = -\frac53 i \inner{a}{\tau} \gamma_5 (N_6 - N_7) + 4 i \gamma_5 \inner{a}{{(\Delta_6 - \Delta_7)}} \, ,
\\ \delta_5^{\vec{a}} {(\Delta_6 - \Delta_7)}^i = \frac23 i \tau^i \gamma_5 \inner{a}{{(\Delta_6 - \Delta_7)}} - i \inner{a}{\tau} \gamma_5 {(\Delta_6 - \Delta_7)}^i
\\ ~~~~~~~~~~~~~~~~~~~~ + \frac43 i\gamma_5 a^j P^{ij}_{3/2} (N_6 - N_7) \, ,
\end{array} \right .
\\ \nonumber && \delta_5^{\vec{a}} {(\Delta_6 + \Delta_7)}^i = - 2 i \tau^i \gamma_5 \inner{a}{{(\Delta_6 - \Delta_7)}} + 3 i \inner{a}{\tau} \gamma_5 {(\Delta_6 - \Delta_7)}^i \, .
\end{eqnarray}
The same is also true for $N_8$, $N_9$, and $\Delta_8^i$:
\begin{eqnarray}
&& \delta_{5}^{\vec{a}} (N_8 - N_9) = i \inner{a}{\tau} \gamma_5 (N_8 - N_9) \, ,
\\ \nonumber && \left \{ \begin{array}{l}
\delta_{5}^{\vec{a}} ({ N_8 } + 3 N_9) = -\frac53 i \inner{a}{\tau} \gamma_5 ({ N_8 } + 3 N_9) - 8 i \gamma_5 \inner{a}{\Delta_8} \, ,
\\ \delta_5^{\vec{a}} \Delta_8^i = - \frac23 i \gamma_5 a^j \Iprot{i}{j} ({ N_8 } + 3 N_9) + \frac23 i \tau^i \gamma_5 \inner{a}{\Delta_8}
\\ ~~~~~~~~~~ - i\inner{a}{\tau}\gamma_5 \Delta_8^i \, .
\end{array} \right .
\end{eqnarray}

\subsection{$J = {3 \over 2}$}

Under the Abelian $U(1)_A$ chiral transformation, the $D(1,\frac12)_{I=\frac12}$ three-quark baryon fields transform as
\begin{eqnarray}
\nonumber && \left \{ \begin{array}{l}
\delta_5 N_{3\mu} = i a \gamma_5 N_{3\mu} \, ,
\\ \delta_5 N_{4\mu} = i a \gamma_5 N_{4\mu} \, ,
\\ \delta_5 N_{5\mu} = i a \gamma_5 N_{5\mu} \, ,
\end{array} \right.
\\ &&
\left \{ \begin{array}{l}
\delta_5 N_{8\mu} = i a \gamma_5 N_{8\mu} \, ,
\\ \delta_5 N_{9\mu} = i a \gamma_5 N_{9\mu} \, ,
\\ \delta_5 N_{10\mu} = i a \gamma_5 N_{10\mu} \, .
\end{array} \right.
\end{eqnarray}
The $D(1,\frac12)_{I=\frac32}$ baryon fields transform as
\begin{eqnarray}
\left \{ \begin{array}{l}
\delta_5 \Delta^i_{4\mu} = i a \gamma_5 \Delta^i_{4\mu} \, ,
\\ \delta_5 \Delta^i_{5\mu} = i a \gamma_5 \Delta^i_{5\mu} \, ,
\\ \delta_5 \Delta^i_{8\mu} = i a \gamma_5 \Delta^i_{8\mu} \, .
\end{array} \right.
\end{eqnarray}
The $D(\frac32,0)_{I=\frac12}$ and $D(\frac32,0)_{I=\frac32}$ baryon fields transform as
\begin{eqnarray}
\left \{ \begin{array}{l}
\delta_5 N_{5\mu\nu} = 3 i a \gamma_5 N_{3\mu\nu} \, ,
\\ \delta_5 N_{10\mu\nu} = 3 i a \gamma_5 N_{10\mu\nu} \, ,
\\ \delta_5 \Delta^i_{5\mu\nu} = 3 i a \gamma_5 \Delta^i_{5\mu\nu} \, .
\end{array} \right.
\end{eqnarray}

Under the $SU(2)_A$ chiral transformation, the $D(1,\frac12)_{I=\frac12}$ baryon fields transform as
\begin{eqnarray}
&& \nonumber \left \{ \begin{array}{l}
\delta_5^{\vec{a}} N_{3\mu} = i \inner{a}{\tau} \gamma_5 N_{3\mu} + \frac23 i \inner{a}{\tau}\gamma_5 N_{4\mu} + 2i\gamma_5\inner{a}{\Delta_4^\mu} \, ,
\\ \delta_5^{\vec{a}} N_{4\mu} = 2 i \inner{a}{\tau} \gamma_5 N_{3\mu} - \frac13 i \inner{a}{\tau}\gamma_5 N_{4\mu} + 2i\gamma_5\inner{a}{\Delta_4^\mu} \, ,
\\ \delta_5^{\vec{a}} N_{5\mu} = \frac53 i\inner{a}{\tau} \gamma_5 N_{5\mu} - 4 i \gamma_5\inner{a}{\Delta_{5\mu}} \, ,
\end{array} \right.
\\ && \nonumber \left \{ \begin{array}{l}
\delta_5^{\vec{a}} N_{8\mu} = - \frac13 i\inner{a}{\tau} \gamma_5 N_{8\mu} + 2 i\inner{a}{\tau} \gamma_5 N_{9\mu} + 2 i \gamma_5 \inner{a}{\Delta_{8\mu}} \, ,
\\ \delta_5^{\vec{a}} N_{9\mu} = \frac23 i\inner{a}{\tau} \gamma_5 N_{8\mu} + i\inner{a}{\tau} \gamma_5 N_{9\mu} + 2 i \gamma_5 \inner{a}{\Delta_{8\mu}} \, ,
\\ \delta_5^{\vec{a}} N_{10\mu} = - i \inner{a}{\tau} \gamma_5 N_{10\mu} \, .
\end{array} \right.
\\
\end{eqnarray}
The $D(1,\frac12)_{I=\frac32}$ baryon fields transform as
\begin{eqnarray}
\left \{ \begin{array}{l}
\delta_5^{\vec{a}} \Delta_4^{\mu i} = 2 i\gamma_5 a^j P^{ij}_{\frac32} N_{3\mu} + \frac23 i\gamma_5 a^j P^{ij}_{\frac32} N_{4\mu}
\\ ~~~~~~~~~~ -\frac23 i \tau^i\gamma_5\inner{a}{\Delta_4^\mu}+i\inner{a}{\tau}\gamma_5 \Delta_4^{\mu i} \, ,
\\ \delta_5^{\vec{a}} \Delta_5^{\mu i} = - \frac43 i\gamma_5 a^j P^{ij}_{3/2} N_5^\mu - \frac23 i \tau^i\gamma_5 \inner{a}{\Delta_5^\mu}
\\ ~~~~~~~~~~ + i\inner{a}{\tau}\gamma_5 \Delta_5^{\mu i} \, ,
\\ \delta_5^{\vec{a}} \Delta_8^{\mu i} = 2 i\gamma_5 a^j P^{ij}_{\frac32} N_{9\mu} + \frac23 i\gamma_5 a^j P^{ij}_{\frac32} N_{8\mu}
\\ ~~~~~~~~~~ -\frac23 i \tau^i\gamma_5\inner{a}{\Delta_8^\mu}+i\inner{a}{\tau}\gamma_5 \Delta_8^{\mu i} \, .
\end{array} \right.
\end{eqnarray}
The $D(\frac32,0)_{I=\frac12}$ and $D(\frac32,0)_{I=\frac32}$ baryon fields transform as
\begin{eqnarray}
\left \{ \begin{array}{l}
\delta_5^{\vec{a}}N_{5\mu\nu} = i\inner{\tau}{a}\gamma_5 N_{5\mu\nu} \, ,
\\ \delta_5^{\vec{a}}N_{10\mu\nu} = i\inner{\tau}{a}\gamma_5 N_{10\mu\nu} \, ,
\\ \delta_5^{\vec{a}} \Delta_{5\mu\nu}^i = - 2 i \tau^i \gamma_5 \inner{a}{\Delta_{5\mu\nu}} + 3 i \inner{a}{\tau} \gamma_5 \Delta_{5\mu\nu}^i \, ,
\end{array} \right.
\end{eqnarray}
We find that $N_{3\mu}$, $N_{4\mu}$, and $\Delta_{4\mu}^i$ can be reduced to irreducible components by taking the following linear combinations:
\begin{eqnarray}
&& \delta_5^{\vec{a}} (N_{3\mu} - N_{4\mu}) = - i \inner{a}{\tau} \gamma_5 (N_{3\mu} - N_{4\mu}) \, ,
\\ && \nonumber \left \{ \begin{array}{l}
\delta_5^{\vec{a}} (3 N_{3\mu} + {N_{4\mu}}) = \frac53 i \inner{a}{\tau} \gamma_5 (3 N_{3\mu} + {N_{4\mu}}) + 8 i\gamma_5\inner{a}{\Delta_4^\mu} \, ,
\\ \delta_5^{\vec{a}} \Delta_4^{\mu i} = \frac23 i\gamma_5 a^j P^{ij}_{\frac32} (3 N_{3\mu} + {N_{4\mu}}) -\frac23 i \tau^i\gamma_5\inner{a}{\Delta_4^\mu}
\\ ~~~~~~~~~~+i\inner{a}{\tau}\gamma_5 \Delta_4^{\mu i} \, .
\end{array} \right.
\end{eqnarray}
The same can be done for $N_{8\mu}$, $N_{9\mu}$, and $\Delta_{8\mu}^i$:
\begin{eqnarray}
&& \delta_5^{\vec{a}} (N_{8\mu} - N_{9\mu}) = - i\inner{a}{\tau} \gamma_5 (N_{8\mu} - N_{9\mu}) \, ,
\\ && \nonumber \left \{ \begin{array}{l}
\delta_5^{\vec{a}} ({N_{8\mu}} + 3 N_{9\mu}) = \frac53 i\inner{a}{\tau} \gamma_5 ({N_{8\mu}} + 3 N_{9\mu}) + 8 i \gamma_5 \inner{a}{\Delta_{8\mu}} \, ,
\\ \delta_5^{\vec{a}} \Delta_8^{\mu i} = \frac23 i\gamma_5 a^j P^{ij}_{\frac32} ({N_{8\mu}} + 3 N_{9\mu}) - \frac23 i \tau^i\gamma_5\inner{a}{\Delta_8^\mu}
\\ ~~~~~~~~~~ +i\inner{a}{\tau}\gamma_5 \Delta_8^{\mu i} \, .
\end{array} \right.
\end{eqnarray}
We note that $N_{5\mu}$ and $\Delta_{5\mu}^i$ are also related to each other under the $SU(2)_A$ chiral transformation, and the linear combinations are not needed.

\section{Conclusions and Summary}
\label{sec:summary}

In this section, we summarize the chiral multiplets found in the previous section. For the $J=\frac12$ baryon fields, $(N_1 \pm N_2)$, $(N_3 - N_4)$, $N_5$, $(N_6 + N_7)$, $(N_8 - N_9)$, and $N_{10}$ form seven $[(\frac12, 0)\oplus(0, \frac12)]$ chiral multiplets; $(N_3 + \frac13 N_4, \Delta^i_4)$, $(N_6 - N_7 ,\Delta^i_6 - \Delta^i_7)$, and $(N_8 + 3 N_9 ,\Delta^i_8)$ form three $[(1, \frac12)\oplus(\frac12, 1)]$ chiral multiplets; $\Delta^i_5$ and $(\Delta^i_6 + \Delta^i_7)$ form two $[(\frac32, 0)\oplus(0, \frac32)]$ chiral multiplets. For the $J = \frac32$ baryon fields, $(N_{3\mu} - N_{4\mu})$, $(N_{8\mu} - N_{9\mu})$, $N_{10\mu}$, $N_{5\mu\nu}$, and $N_{10\mu\nu}$ form five $[(\frac12, 0)\oplus(0, \frac12)]$ chiral multiplets, $(N_{3\mu} + \frac13 N_{4\mu}, \Delta^i_{4\mu})$, $(N_{5\mu}, \Delta^i_{5\mu})$, and $(N_{8\mu} + 3 N_{9\mu}, \Delta^i_{8\mu})$ form three $[(1, \frac12)\oplus(\frac12, 1)]$ chiral multiplets; $\Delta^i_{5\mu\nu}$ forms one $[(\frac32, 0)\oplus(0, \frac32)]$ chiral multiplet.

We find a total of 31 independent chiral multiplets. They are listed in Tables~\ref{tab:spin12},~\ref{tab:spin32a}, and \ref{tab:spin32b} with their Abelian axial charge $g_A^{(0)}$ and the non-Abelian axial charge $g_A^{(1)}$. The $SU(2)_A$ chiral transformation of $\Delta^i$ fields always contain two diagonal terms: $\tau^i \inner{a}{\Delta}$ and $\inner{a}{\tau} \Delta^i$. To show ``numbers'' in these tables, we simply add their coefficients and use brackets to denote this, such as ``$g_A^{(1)} = (-\frac13)$'' for the $\Delta_4^i$ field. We note that some of the fields have already been studied in the bi-local case~\cite{Dmitrasinovic:2011yf}.

We find that the chiral transformation properties of every new field, which turns up in the tri-local case, is similar to some old one in the local case. This is also true for the bi-local baryon fields, i.e., the chiral transformation properties of every new field, which turns up in the bi-local case, is similar to some old one in the local case. We find that the $[(\frac12, 0)\oplus(0, \frac12)]$ chiral multiplets $(N_1 - N_2)$, $(N_3 - N_4)$, and $(N_8 - N_9)$ transform in the same way, whereas the $[(0, \frac12)\oplus(\frac12, 0)]$ chiral multiplets $(N_{3\mu} - N_{4\mu})$, $(N_{8\mu} - N_{9\mu})$, and $N_{10\mu\nu}$ transform like their mirror fields; the $[(\frac12, 0)\oplus(0, \frac12)]$ chiral multiplets $(N_1 + N_2)$, $N_5$, $(N_6 + N_7)$, $N_{10}$, $N_{5\mu\nu}$, and $N_{10\mu\nu}$ transform in the same way; the $[(\frac12,1)\oplus(1, \frac12)]$ chiral multiplets $(N_3 + \frac13 N_4, \Delta^i_4)$, $(N_6 - N_7 ,\Delta^i_6 - \Delta^i_7)$, and $(N_8 + 3 N_9 ,\Delta^i_8)$ transform in the same way, whereas the $[(1, \frac12)\oplus(\frac12,1)]$ chiral multiplets $(N_{3\mu} + \frac13 N_{4\mu}, \Delta^i_{4\mu})$, $(N_{5\mu}, \Delta^i_{5\mu})$, and $(N_{8\mu} + 3 N_{9\mu}, \Delta^i_{8\mu})$ transform like their mirror fields; the $[(\frac32, 0)\oplus(0, \frac32)]$ chiral multiplets $\Delta_5^i$, $(\Delta_6^i + \Delta_7^i)$, and $\Delta_{5}^{\mu \nu}$ transform in the same way.

In summary, we have systematically investigated tri-local three-quark baryon fields. Through the Fierz transformation, we find that $B(x,y,z)$, $B(y,z,x)$, and $B(z,x,y)$ can be related using some transformation matrices (others like $B(y,x,z)$ can be also related, see Eq.~(\ref{eq:relation})). The Pauli principle is taken into account also by using the Fierz transformation. However, different from the local and bi-local cases, in the case of tri-local baryon fields, the Pauli principle does not forbid any possible structure among the three quark fields inside, whose structure can be carried out by suitable Lorentz and flavor matrices.

We classified all the $J=\frac12$ and $J=\frac32$ baryon fields and studied their chiral transformation properties, according to their Lorentz and isospin (flavor) group representations. Together with the local and bi-local three-quark baryon fields, which have been studied in Refs.~\cite{Nagata:2007di,Dmitrasinovic:2011yf}, we arrive at our final conclusion that the axial coupling constant $|g_A| = \frac{5}{3}$ only exists for nucleon fields belonging to the chiral representation $(\frac12,1) \oplus (1,\frac12)$ where the nucleon and $\Delta$ fields are both inside. Moreover, all the nucleon fields belonging to this chiral representation have $|g_A| = \frac{5}{3}$.

The procedures in this paper can be straightforwardly applied to bi-local/tri-local three-quark baryon fields having $SU(3)_L \otimes SU(3)_R$ chiral symmetry, which is our next subject.

\begin{table}[tbh]
\begin{center}
\caption{Chiral multiplets of Lorentz representation $D(\frac{1}{2},0)$, together with their Abelian and non-Abelian axial charges $g_A^{(0)}$ and $g_A^{(1)}$.}
\begin{tabular}{cccc}
\hline \hline
\mbox{Baryon Fields}& $g_A^{(0)}$ & $g_A^{(1)}$ & $SU_L(2) \times SU_R(2)$
\\ \hline \hline
$N_1 - N_2$ & $-1$ & $+1$ & $(\frac12,0) \oplus (0,\frac12)$
\\ \hline
$N_1 + N_2$ & $+3$ & $+1$ & $(\frac12,0) \oplus (0,\frac12)$
\\ \hline
$N_3 - N_4$ & $-1$ & $+1$  & $(\frac12,0) \oplus (0,\frac12)$
\\ \hline
$N_5$ & $+3$& $+1$ & $(\frac12,0) \oplus (0,\frac12)$
\\ \hline
$N_6 + N_7$ & $+3$ & $+1$ & $(\frac12,0) \oplus (0,\frac12)$
\\ \hline
$N_8 - N_9$ & $-1$ & $+1$ & $(\frac12,0) \oplus (0,\frac12)$
\\ \hline
$N_{10}$ & $+3$& $+1$ & $(\frac12,0) \oplus (0,\frac12)$
\\ \hline \hline
$N_3 + \frac13 N_4$ & $-1$& $-\frac53$ & $(\frac12,1) \oplus (1,\frac12)$
\\
$\Delta_4^i$ & $-1$& $(-\frac13)$ & $(\frac12,1) \oplus (1,\frac12)$
\\ \hline
$N_6 - N_7$ & $-1$ & $-\frac53$ & $(\frac12,1) \oplus (1,\frac12)$
\\
$\Delta_6^i - \Delta_7^i$ & $-1$& $(-\frac13)$ & $(\frac12,1) \oplus (1,\frac12)$
\\ \hline
$N_8 + 3 N_9$ & $-1$ & $-\frac53$ & $(\frac12,1) \oplus (1,\frac12)$
\\
$\Delta_8^i$ & $-1$& $(-\frac13)$ & $(\frac12,1) \oplus (1,\frac12)$
\\ \hline \hline
$\Delta_5^i$ & $+3$& $(+1)$ & $(\frac32,0) \oplus (0,\frac32)$
\\ \hline
$\Delta_6^i + \Delta_7^i$ & $+3$& $(+1)$ & $(\frac32,0) \oplus (0,\frac32)$
\\ \hline \hline
\end{tabular}
\label{tab:spin12}
\end{center}
\end{table}
\begin{table}[tbh]
\begin{center}
\caption{Chiral multiplets of Lorentz representation $D(1,\frac{1}{2})$, together with their Abelian and non-Abelian axial charges $g_A^{(0)}$ and $g_A^{(1)}$.}
\begin{tabular}{ccccc}
\hline \hline
\mbox{Baryon Fields}& $g_A^{(0)}$ & $g_A^{(1)}$ & $SU_L(2) \times SU_R(2)$
\\ \hline \hline
$N_3^{\mu} - N_4^{\mu}$ & $+1$ & $-1$  & $(0,\frac12) \oplus (\frac12,0)$
\\ \hline
$N_8^{\mu} - N_9^{\mu}$ & $+1$ & $-1$  & $(0,\frac12) \oplus (\frac12,0)$
\\ \hline
$N_{10}^{\mu}$ & $+1$ & $-1$  & $(0,\frac12) \oplus (\frac12,0)$
\\ \hline \hline
$N_3^{\mu} + \frac13 N_4^{\mu}$ & $+1$& $+\frac53$ & $(1,\frac12) \oplus (\frac12,1)$
\\
$\Delta_4^{\mu i}$ & $+1$& $(+\frac13)$ & $(1,\frac12) \oplus (\frac12,1)$
\\ \hline
$N_5^{\mu}$ & $+1$ & $+\frac53$ & $(1,\frac12) \oplus (\frac12,1)$
\\
$\Delta_5^{\mu}$ & $+1$ & $(+\frac13)$ & $(1,\frac12) \oplus (\frac12,1)$
\\ \hline
$N_8^{\mu} + 3 N_9^{\mu}$ & $+1$& $+\frac53$ & $(1,\frac12) \oplus (\frac12,1)$
\\
$\Delta_8^{\mu i}$ & $+1$& $(+\frac13)$ & $(1,\frac12) \oplus (\frac12,1)$
\\ \hline \hline
\end{tabular}
\label{tab:spin32a}
\end{center}
\end{table}
\begin{table}[tbh]
\begin{center}
\caption{Chiral multiplets of Lorentz representation $D(\frac{3}{2},0)$, together with their Abelian and non-Abelian axial charges $g_A^{(0)}$ and $g_A^{(1)}$.}
\begin{tabular}{cccc}
\hline \hline
& $U_A(1)$ & $SU_A(2)$ & $SU_V(2) \times SU_A(2)$
\\ \hline \hline
$N_{5}^{\mu \nu}$ & $+3$ & $+1$ & $(\frac12,0) \oplus (0,\frac12)$
\\ \hline
$N_{10}^{\mu \nu}$ & $+3$ & $+1$ & $(\frac12,0) \oplus (0,\frac12)$
\\ \hline
$\Delta_{5}^{\mu \nu}$ & $+3$ & $(+1)$ & $(\frac32,0) \oplus (0,\frac32)$
\\ \hline \hline
\end{tabular}
\label{tab:spin32b}
\end{center}
\end{table}

\section*{Acknowledgments}

This work is partly supported by the National Natural Science Foundation of China under Grant No. 11147140, and the Scientific Research Foundation for the Returned Overseas Chinese Scholars, State Education Ministry.

%

\end{document}